\documentclass[aps,prl,superscriptaddress]{revtex4}
\usepackage{graphicx}

\begin{document}


\title{Funneling Light Through a Subwavelength Aperture with Epsilon-Near-Zero Materials}


\author{David Slocum}
\author{Sandeep Inampudi}
\author{David C. Adams}
\author{Shivashankar Vangala}
\affiliation{University of Massachusetts Lowell, Lowell, MA 01854}
\author{Nicholas A. Kuhta}
\affiliation{Oregon State University, Corvallis, OR 97331}
\author{William D. Goodhue}
\author{Viktor A. Podolskiy}
\author{Daniel Wasserman}
\affiliation{University of Massachusetts Lowell, Lowell, MA 01854}

\date{\today}



\maketitle



{\bf
Integration of the next generation of photonic structures with electronic and optical on-chip components requires the development of effective methods for confining and controlling light in subwavelength volumes. Several techniques enabling light coupling to sub-wavelength objects have recently been proposed, including grating- \cite{Merlin,Eleftheriades,Podolskiy1}, and composite-based solutions\cite{superlens,Zhang1,hyperlensEN,hyperlensNE,hyperlensIS}. However, experimental realization of these couplers involves complex fabrication with $\sim 10nm$ resolution in three dimensions. One promising alternative to complex coupling structures involves materials with vanishingly small dielectric permittivity, also known as epsilon-near-zero (ENZ) materials\cite{Ziol,Engheta1}. In contrast to the previously referenced approaches, a single flat layer of ENZ-material is expected to provide efficient coupling between free-space radiation and sub-wavelength guiding structures. Here we report the first direct observation of bulk-ENZ-enhanced transmission through a subwavelength slit, accompanied by a theoretical study of this phenomenon. Our study opens the door to multiple practical applications of ENZ materials and ENZ-based photonic systems.
}

\begin{figure}
\includegraphics[width=6cm]{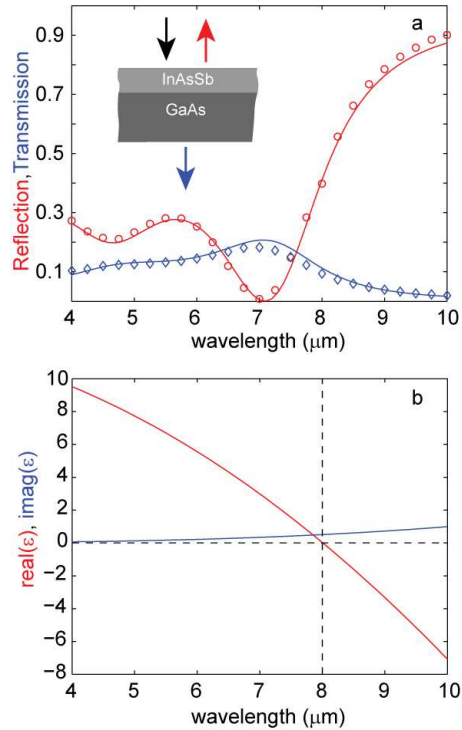}%
 \caption{\label{Figure1} {\bf Bulk properties of ENZ material}; Panel (a) provides a schematic of the optical measurements on the bulk ENZ material, and the related transmission and reflection data from these measurements; symbols show experimental reflection (red circles) and transmission (blue diamonds); solid lines represent theoretical fits (see Methods); panel (b) shows retrieved permittivity spectra.
 }
 \end{figure}

It has been known that ENZ materials possess some unusual properties \cite{Ziol}. In particular, the electromagnetic field inside the ENZ structures tends to become completely homogeneous, reflecting the dramatic extension of local wavelength caused by a vanishingly small refractive index. More recently, it has been shown theoretically \cite{Engheta1} that lossless ENZ systems enhance coupling between two planar waveguides through an ultra-thin guiding channel.  Experimental verification of this principle, in the microwave regime, was performed in systems where the ENZ coupler utilized split ring resonator structures \cite{DSmith1} or alternatively, is replaced with a waveguide designed to mimic the optical properties of an ENZ material \cite{Engheta2}. While the original waveguide coupling structures discussed in the literature are not directly applicable to macroscopic plane-wave-to-nanoscale couplers, an alternative mechanism for ENZ-enhanced transmission of plane waves through a relatively thick ENZ cover, discussed using a ray optics ``focusing'' formalism, was presented in \cite{Engheta3}.  However, to the best of our knowledge, up to now no experiments with homogenous bulk ENZ materials have been performed. In this work we present the first experimental study of ENZ-enhanced transmission through subwavelength features with a bulk ENZ material.  Our material is semiconductor-based and inclusion-free, and operates at optical frequencies (in the mid-infrared wavelength range), opening the door to integration of ENZ materials with cutting edge optoelectronic materials and devices.  We also present a comprehensive analysis of this unique phenomenon and discuss the relationship between two ENZ-enhanced transmission geometries.

The ENZ material used in this experiment consists of the heavily doped, narrow band-gap semiconductor, InAs$_x$Sb$_{1-x}$, with $x=0.89$.  Our InAsSb layer was grown by Molecular Beam Epitaxy (MBE) in a Riber 32 system, on a semi-insulating GaAs substrate.  The InAsSb epilayer is 1.15$\mu m$ thick and heavily n-doped ($\sim1-2\times10^{19}cm^{-3}$).  It is well known that conduction band electrons in semiconductors can behave as a free-electron plasma, whose interaction with incident radiation fields can be described by a Drude-like dispersion model.  In such a system, the permittivity of the semiconductor can be described using:
\begin{equation}
\label{epsdrude}
\epsilon(\omega)=\epsilon_\infty\left(1-\frac{\omega_p^2}{\omega^2+i\omega\Gamma}\right)   ,   \omega_p^2=\frac{ne^2}{m^*\epsilon_o\epsilon_\infty},
\end{equation}
where $\omega$ is the radial frequency of the radiation, $\Gamma$ is the damping constant, $\epsilon_\infty$ is the high-frequency permittivity of the semiconductor lattice, $\omega_p$ is the semiconductor plasma frequency, $n$ is the carrier concentration, and $m^*$ the effective mass of the free electrons in the semiconductor.  By using the InAsSb material system, which can be doped heavily n-type, and which has a low conduction band electron effective mass, we can push the plasma frequency of the material well into the $6-10\mu m$ wavelength range.

\begin{figure}
\includegraphics[width=10cm]{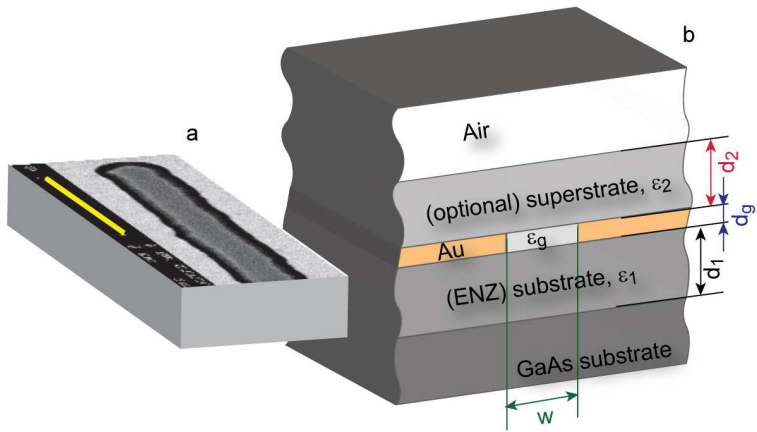}%
 \caption{\label{figDesign} {\bf Geometry of the system}; (a) SEM image of a representative ENZ-coupled subwavelength aperture used in experiments (length of the yellow scale bar is $6 \mu m$). (b) schematic drawing of the structure geometry.  The total thickness of the ENZ material, n-doped InAsSb, is $d_1+d_g\simeq 1.15\mu m$.  The thickness of both the gold (and the slit) is $d_g\simeq 300nm$.  The slit width ($w$), for the samples studied, varied from $1.0\mu m$ to $1.8\mu m$ in $0.2\mu m$ increments. }
 \end{figure}

The bulk optical properties of our InAsSb layer have been derived from optical measurements as described in Methods, and demonstrate that the material exhibits ENZ behavior in the vicinity of $\lambda_0=8\mu m$.  The resultant spectral dependence of the real and imaginary parts of the InAsSb permittivity are shown in Fig.\ref{Figure1}.  Values of $\epsilon^{''}$ for our material are of the same order as $\epsilon^{''}$ for noble metals (Ag, Au) or more exotic materials such as transparent conducting oxides \cite{Bolt} about their plasma frequencies in the near UV and near-IR wavelength ranges, respectively. Therefore, our studies provide an outlook for the performance of realistic plasmonic materials across the optical frequency ranges, from UV to mid-IR. In order to experimentally demonstrate ENZ-enhanced transmission through subwavelength optical features, subwavelength slits in an optically thick metal layer were fabricated onto the InAsSb epilayer, as shown in Fig.\ref{figDesign} (see Methods). At the same time, a set of control samples, with  slit widths identical to the InAsSb epilayer samples, were patterned on a semi-insulating GaAs substrate, in order to compare the ENZ-enhanced transmission to structures with identical geometries, but without material losses. The GaAs substrate, having relatively large refractive index ($n\simeq 3.3$) represent the best-case scenario for traditional direct coupling of light into the small slit.

\begin{figure}
 \includegraphics[width=12cm]{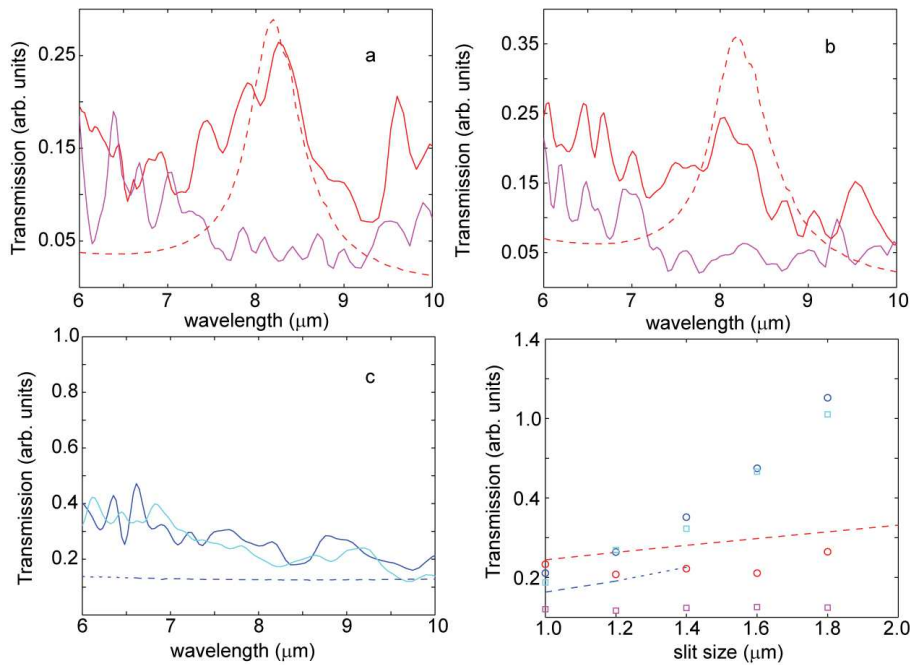}%
 \caption{\label{Figure2} {\bf Analysis of ENZ-enhanced transmission}; The four panels represent the analysis of transmission through ENZ-enhanced and control samples; Panels (a,b) show transmission spectra for TM (red) and TE (magenta) polarized light, for the systems with (a) $w=1\mu m$ and (b) $w=1.4 \mu m$; Panel (c) illustrate the transmission through the control structure where GaAs replaces ENZ material ($w=1 \mu m$); Panel (d) shows the transmitted power at the ENZ-enhanced transmission peak $\lambda \sim 8 \mu m$ as a function of slit width, and demonstrates one of the essential characteristics of ENZ-enhanced tranmission: its extremely weak dependence on the slit size $w$. In all panels, symbols and solid lines represent experimental data, dashed lines represent the results from the analytical technique developed in this work; The relatively poor agreement between the analytical results and experimental transmission through GaAs control samples is explained by relatively large effective slit width ($nw$) in these systems.}
 \end{figure}

The transmission spectra of the ENZ-based and the control structures are shown in Fig.\ref{Figure2}(a-c).  It can clearly be seen that the InAsSb samples provide a drastic enhancement for TM-polarized light coupling into the slit structure at or near the ENZ wavelength.  In fact, for systems with narrower slits ($\sim 1\mu m$), a higher transmission intensity predicted (and experimentally observed) for the ENZ samples (at $\lambda_{0}\sim 8\mu m$) then their dielectric high-permittivity control counterparts, despite the significant losses associated with free carriers in the ENZ samples. Note that the ENZ material does not provide transmission enhancement for TE-polarized radiation.

To understand the origin of the observed phenomenon, transmission through the system has been modeled with mode-matching software, developed according to Ref.\cite{sukoModeMatch}, followed by the development of an analytical technique describing transmission and reflection by the system, outlined in Methods. These studies clearly demonstrate that transmission through the system is dominated by the fundamental ($k_x=0$) mode of the planar guide formed by the slit. Since this mode only exists for TM-polarized light, TE waves do not exhibit ENZ-related transmission peaks.

The amplitude of the fundamental mode, and correspondingly, the total light transfer through the system, is related to the interplay between two processes: (1) coupling between the incident radiation and the $k_x=0$ mode at the entrance slit of the device; and (2) propagation of the mode through the slit. The flat ENZ layer at the entrance side of the device drastically enhances the coupling between free-space and the slit (guide). This enhancement is related to amplification of the amplitude of the electromagnetic field inside the ENZ material \cite{He}. In our system, this enhancement originates from bulk plasma resonance of the doped semiconductor.

\begin{figure}
 \includegraphics[width=8cm]{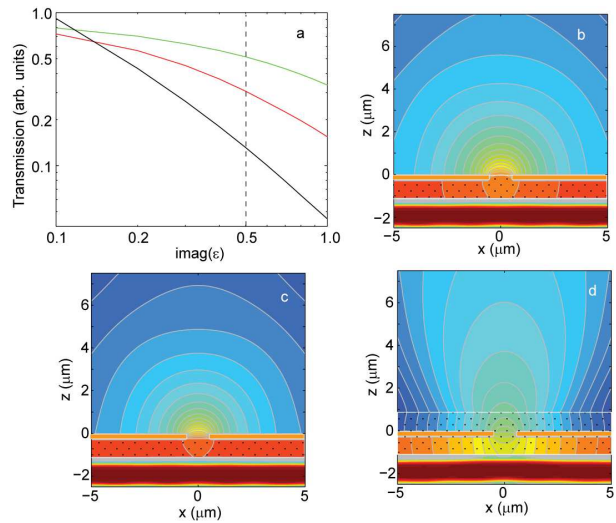}%
 \caption{\label{figLowLoss} {\bf ENZ-assisted light funneling at different losses} Panel (a) represents the comparison between transmitted power at the ENZ wavelength in three ENZ-assisted structures: (i) the system realized in our experiments (red), when ENZ material fills the substrate and in-slit regions, (ii) the system where the ENZ material is deposited as a thin substrate layer, but air fills both the slit region and the region above the slit (green), and finally, (iii) the design originally suggested in Ref.\cite{Engheta3} where ENZ fills substrate, slit, and superstrate regions of the geometry shown in Fig.\ref{figDesign} (black). It is clearly seen that at relatively large absorption, the winning design calls for low-loss material (air) inside the guide and above it; the practical benefits of ENZ are limited to extending the wavelength inside the system. Lower-loss systems, however, benefit from the extra ENZ layers inside and behind the slit; In the limit of low loss, the ENZ light collimation, suggested in Ref.\cite{Engheta3} is seen in the structure with the ENZ superstrate. Panels (b-d) illustrate field distribution in the three systems (i-iii, respectively) at $imag(\epsilon)=0.1$. }
 \end{figure}

The secondary enhancement comes about from modal inter-coupling between the ENZ-filled guide and the bulk ENZ coupling layer. However, the extra in-coupling efficiency, provided by the ENZ- layer inside the waveguide is more than compensated for by the negative effect of the ENZ-related loss of the device as the mode propagates through the system, which ultimately reduces the amplitude of the fundamental mode at the exit side of the slit. Since the fabrication of ENZ-filled systems is significantly more complicated than the relatively straightforward fabrication of subwavelength geometries above planar ENZ materials, the benefits of complex systems have to be weighed against their expected performance. Such a comparison is shown in Fig.\ref{figLowLoss}.  It is seen that for realistic values of loss, transmission through the subwavelength channel is maximized for structures where the only ENZ component is the thin coupling layer, and the slit is free of ENZ material. When losses are substantially reduced, however, the positive in-coupling boost provided by the ENZ material inside the slit, accompanied by an ENZ out-coupling layer, add performance to the system.

To conclude, we have reported the first comprehensive study of ENZ-enhanced light transmission through an isolated subwavelength slit. An analytical model describing the transmission has been developed, and the interplay between different contributions to ENZ-enhancement have been analyzed. Theoretical results are in agreement with experimental data. The developed formalism opens the door for ENZ-enhanced coupling to a variety of subwavelength plasmonic, beamsteering, nanophotonic, and nanoelectronic \cite{EOT,beamsteer,superlens,hyperlensEN,hyperlensNE,cloakJP,cloakVS,blackHoleVS,enghetaCircuits} systems across a broad range of optical frequencies.

 The authors gratefully acknowledge N. Engheta (UPenn) for useful discussion and comments.  The SEM image in Fig. \ref{figDesign} was taken by Dave Safford at IES Technical Sales with a PhenomWorld SEM model PHENOM.  This work was supported by the AFOSR Young Investigator Program under Grant \#FA9550-10-1-0226 and NSF Grant \# ECCS-0724763.

\section{Methods}

{\bf Permittivity of doped semiconductors} The bulk optical properties of our ENZ layer can be determined from the fitting process outlined in Ref. \cite{Li}, specifically developed for a narrow band semiconductor (InAs) deposited upon GaAs.  Transmission and reflection data were collected on the as-grown InAsSb/GaAs wafer, and this data fitted by adjusting the plasma frequency $\omega_p$ and damping constant $\Gamma$ until excellent agreement between modeled and experimental data was obtained.  From this, the plasma wavelength of the material was determined to be $\sim8\mu m$ ($\sim1250cm^{-1}$) and the damping constant $\Gamma=50cm^{-1}$, in good agreement with the results of Ref. \cite{Li}.  It should be noted that in Fig. \ref{Figure2} the plasma frequency does not, in fact, correspond to the minimum (maximum) of the reflected (transmitted) signal, as this data is a convolution of the InAsSb material properties and the Fabry-Perot cavity formed by the InAsSb layer on the GaAs substrate.  In addition, the theoretical transmission data depicted in Fig. \ref{Figure2} has been scaled by a loss factor $\exp(-2\pi l_0/\lambda)$, with $l_0\simeq1.1 \mu m$ that describes absorption from lattice-mismatch-induced defects in the transition layer between the GaAs substrate and the ENZ layer. Alternative models of permittivity that do not involve losses in this transitional layer yield qualitatively incorrect results (see supporting online documentation)

{\bf Device Fabrication.}  The device was prepared using standard lithographic techniques.  Five columns of slits, with slit widths ranging from 1$\mu m$ to 1.8$\mu m$ in 0.2$\mu m$ increments, were patterned onto the semiconductor substrate as a photoresist etch mask.  The samples were then etched 300nm using a citric acid:hydrogen peroxide etchant.  The photoresist etch mask is retained, and the sample coated in 295nm of Au, following a 5nm Ti adhesion layer, after which the slit feature is lifted off.  Both InAsSb epilayer samples and undoped GaAs samples were fabricated simultaneously, so that transmission through ENZ and non-ENZ slit structures could be compared.

{\bf Device Characterization} Slit transmsission data was collected on a Bruker IR1 infrared microscope coupled to a Bruker V70 Fourier Transform Infrared (FTIR) Spectrometer, with a spectral resolution of $32cm^{-1}$.  The internal FTIR globar source acts as the incident radiation source, and a wire-grid polarizer, placed in the incident beam path, directly before the sample, allows us to alternate between TM (electric field perpendicular to the slit) and TE (electric field parallel to the slit) polarized incident light.  The microscope allows seamless switching between optical microscopy and FTIR spectroscopy, allowing for accurate alignment of the detection optics over the subwavelength slit, which is clearly discernable in the visible optical microscope.  Ultimately, the transmitted radiation is collected by a liquid nitrogen cooled HgCdTe (MCT) detector, with a 16$\mu m$ cut-off wavelength.  Data was collected for a spectral range from $700cm^{-1}$ to  $4000cm^{-1}$. Figure \ref{Figure2} shows the TE and TM polarized transmission for $1\mu m$ and $1.4\mu m$ slits fabricated on both the InAsSb and GaAs material.  All transmission data was normalized to transmission through an unpatterned GaAs substrate to remove the effects of substrate absorption, as well as the spectral shape of the emission of the mid-IR internal FTIR source.  A clear transmission peak is seen at ENZ for all TM polarized data on the InAsSb samples, with no such peak observed for the GaAs samples, in either polarization.  Comparing the InAsSb slit transmission spectra to the unpatterned InAsSb transmission spectra shown in Fig. \ref{Figure1}, it is clear that the ENZ peak is not related to any resonant transmission peaks in the bulk InAsSb material.

{\bf Mode-matching calculations} provide unprecedented insight into the optical properties of complex structures. In this approach, the system under investigation is separated into a set of bulk layers and a waveguide layer. The field inside bulk layers is represented as linear combination of plane waves (free-space modes), parametrized by $x$-component of their wavevector. The field inside the slit (waveguide) region is represented as a sum of guided (waveguide) modes with their field concentrated inside the slit, accompanied by a set of free-space-like bulk modes with the field extending throughout the metallic cladding. The spectrum of the bulk modes in the waveguide region of the structure matches the spectrum of the plane waves in the free-space-like regions.

Continuity of the $E_x$ and $H_y$ components of the electromagnetic fields is used to deduce the set of boundary conditions relating the amplitudes of the modes in the adjacent layers. The implementation of the mode-matching simulations used by our group is described in Ref.\cite{sukoModeMatch}. The comparison between mode matching simulations and other numerical techniques for solving Maxwell equations is presented in Supporting Online Information.

{\bf Analytical calculations of light transmission through the system}. To calculate transmission through a subwavelength slit, we represent the incident field as a plane wave with a fixed wavevector $k_x^0$; ($k_x^0=0$ is used in the calculations), and a given amplitude $I_0$; and approximate the reflected field as a combination of the plane wave with the same wavevector and amplitude $R_0$, and a scattered wave with electric field represented as $E_x=S\int{e^{i \vec{k}\cdot\vec{r}} {\rm sinc}(w k_x/2) dk_x}$; further, the transmitted field is approximated as $E_x=T\int{e^{i \vec{k}\cdot\vec{r}} {\rm sinc}(w k_x/2) dk_x}$  (the spectral shape of the reflected and transmitted fields represent the field diffracted by a single slit in the perfect metal conductor and are consistent with results of our numerical calculations). Finally, we use the continuity of $E_x$ and $H_y$ along the slit entrance and impedance boundary conditions\cite{bornWolf} along the metal surface to deduce the values of the parameters $R_0, S, T$, and the amplitudes of the fundamental guided mode $A_g^\pm$ inside the slit in terms of the parameter $I_0$.

The overlap-integral formalism, outlined in Refs.\cite{schevchenkoBook,sukoModeMatch,SingleSlit}, yields the following relationship between the parameters:
\begin{eqnarray}
\label{eqAnalyt}
\left\{
\begin{array}{l}
I_0+R_0+S \frac{2\pi}{w}=  A_g^++A_g^- \Phi_g
\\
\frac{\epsilon_1}{k_1^0}[I_0-R_0-S\cdot I_h(w,\epsilon_1)]=\frac{\epsilon_g}{k_g^0}[A_g^+-A_g^-\Phi_g]
\\
R_0=I_0\rho+S\cdot I_s(w,\epsilon_1,\epsilon_m,q)
\\
\frac{2\pi}{w} T=  A_g^+\Phi_g+A_g^-
\\
\frac{\epsilon_2}{k_2^0}T\cdot I_h(w,\epsilon_2)=\frac{\epsilon_g}{k_g^0}[A_g^+\Phi_g-A_g^-]
\end{array}
\right.
\end{eqnarray}
where $\epsilon_1,\epsilon_g,\epsilon_m, \epsilon_2$ describe permittivities of materials in front, inside, around, and behind the slit (Fig.\ref{figDesign}), $k_\alpha(k_x)=\sqrt{\epsilon_\alpha \omega^2/c^2-{k_x}^2}$, $k_\alpha^0=k_\alpha(k_x^0)$, $\Phi_g=\exp(i k_g^0 d_g)$, Fresnel coefficient $\rho=\frac{\epsilon_1 k_m^0-\epsilon_m k_1^0}{\epsilon_1 k_m^0-\epsilon_m k_1^0}$, and the integral relationships are given by:
\begin{eqnarray}
I_h(w,\epsilon_\alpha)&=&k_\alpha^0\int_{-\infty}^\infty\frac{{\rm sinc}^2({k_x w}/{2})}{k_\alpha(k_x)} dk_x
\\
I_s(w,\epsilon_\alpha,\epsilon_\beta,q)&=&\frac{k_\alpha k_\beta q}{\epsilon_\alpha k_\beta^0+\epsilon_\beta k_\alpha^0}
\int_{-\infty}^\infty \frac{\epsilon_\alpha k_\beta(k_x)+\epsilon_\beta k_\alpha(k_x)}{k_\alpha(k_x) k_\beta(k_x)}
\cdot\frac{k_x \sin(k_x w/2)-q \cos(k_x w/2)}{k_x^2+q^2} dk_x
\nonumber
\end{eqnarray}
The parameter $q\ll 1$ plays the role of the fitting parameter, fine-tuning the role of the impedance boundary conditions. Our studies indicate that the numerical values of the results only weakly depend on the value of $q$ (see Supporting Online Materials). Here we use $q=0.1$.

To account for the effects of the other interfaces (GaAs/ENZ, superstrate/air) either above or below the subwavelength feature in the systems analyzed, Eqs.(\ref{eqAnalyt}) are incorporated into a conventional plane-wave transfer-matrix formalism\cite{bornWolf}.

\section{Supporting online materials}

\begin{figure}
 \includegraphics[width=15cm]{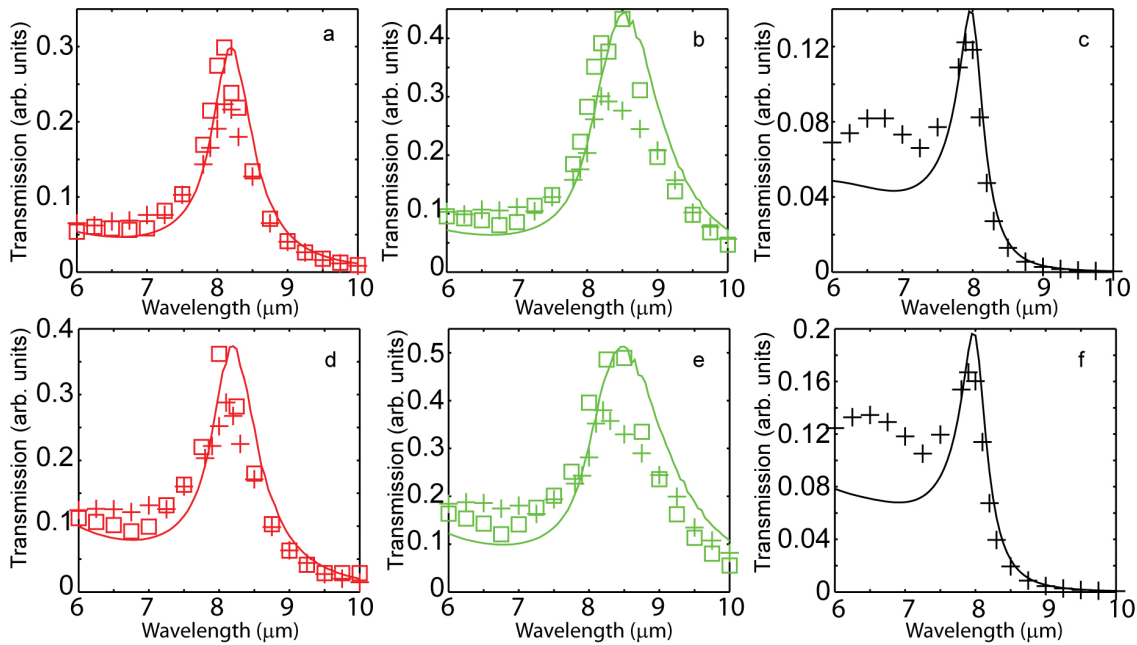}%
 \caption{\label{Figure6} {\bf ENZ-Accuracy of numerical simulations and analytical resuls}. The figure shows the agreement between the light transmission in the three ENZ-assisted structures described in Fig.\ref{figLowLoss}. Crosses, squares, and solid lines correspond to results from FEM simulations, mode-matching solutions of Maxwell equations, and our analytical formalism (Eq.(\ref{eqAnalyt})), respectively; Colors are identical to Fig.\ref{figLowLoss}. The results of all FEM models are scaled by the same number to match the results of mode-matching and analytical calculations, which are plotted with no scaling.
}
\end{figure}

{\bf Accuracy of Numerical Simulations}
The accuracy and stability of our mode-matching simulations were tested for convergence with respect to variations in spectrum of bulk modes, as well as the number of guided modes included in the simulations. Once optimal spectral representation was identified, results of mode-matching solutions of Maxwell equations were compared against the results of a commercial finite-element solver \cite{comsol}. The agreement between the transmission spectra calculated using the two techniques are shown in Fig.\ref{Figure6}

{\bf Alternative models for permittivity, and scaling of analytical data}

As explained in Methods, the Drude model tends to overestimate the transmission through the as-grown ENZ material, indicating the presence of additional losses in the ENZ structures. The origin of these losses can be two-fold. First, because the lattice constant of InAsSb does not match that of the GaAs substrate, there will be a transitional layer at the interface between the two materials that may result in additional transmission losses in ENZ structures.

\begin{figure}
 \includegraphics[width=15cm]{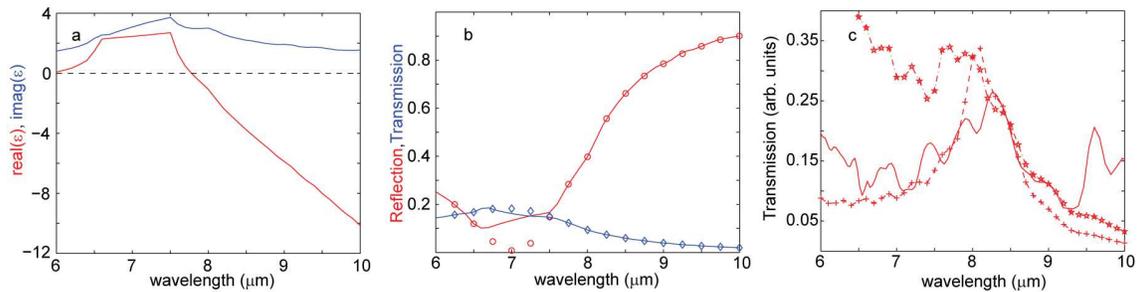}%
 \caption{\label{figEpsComp} {\bf Alternative models of ENZ permittivity}. Panel (a) shows permittivity of InAsSb, extracted under the assumption that additional loss is uniformly distributed through the structure; panel (b) shows theoretical (lines) and experimental (symbols) transmission and reflection spectra of a homogeneous layer of ENZ on GaAs substrate. Finally, panel (c) shows the comparison between ENZ-assisted transmission through $w=1\mu m$ slit, measured experimentally (solid line), and calculated with Drude-based model (crosses) and with the permittivity model shown in panel (a) (stars).
}
\end{figure}

Alternatively, the ``extra" losses could be due to impurities in the ENZ material and thus could be uniformly distributed through the ENZ layers, causing significant deviation of the dispersion of this layer from Drude model predictions. To assess the applicability of these two competing models, we have used a nonlinear fitting approach to extract the permittivity of InAsSb layer directly from tranmission/reflection measurements (not relying on Drude model). The results of this procedure are presented in Fig.\ref{figEpsComp}(a,b).

Furthermore, we calculated the predicted transmission spectra according to the two models of permittivity and compared the results to the experimental data. These calculations are shown in Fig.\ref{figEpsComp}(c). It is clearly seen that only Drude-based model is consistent with experimental results. We therefore conclude that the extra loss seen in our system is localized in a thin transitional layer between InAsSb and GaAs materials. In accordance with this model of our material, the results of our analytical calculations for light transmission through the systems with ENZ substrates in Figs.\ref{Figure1}$-$\ref{figLowLoss} are scaled by the loss factor  $\exp(-2\pi l_0/\lambda)$, with $l_0\simeq1.1 \mu m$.

\bibliography{ENZPRL}

\begin{thebibliography}{9}
\expandafter\ifx\csname natexlab\endcsname\relax\def\natexlab#1{#1}\fi
\expandafter\ifx\csname bibnamefont\endcsname\relax
  \def\bibnamefont#1{#1}\fi
\expandafter\ifx\csname bibfnamefont\endcsname\relax
  \def\bibfnamefont#1{#1}\fi
\expandafter\ifx\csname citenamefont\endcsname\relax
  \def\citenamefont#1{#1}\fi
\expandafter\ifx\csname url\endcsname\relax
  \def\url#1{\texttt{#1}}\fi
\expandafter\ifx\csname urlprefix\endcsname\relax\def\urlprefix{URL }\fi
\providecommand{\bibinfo}[2]{#2}
\providecommand{\eprint}[2][]{\url{#2}}

\bibitem[{\citenamefont{Ziolkowski}(2004)}]{Ziol}
\bibinfo{author}{\bibfnamefont{R.}~\bibnamefont{Ziolkowski}},
  \bibinfo{journal}{Phys. Rev. E} \textbf{\bibinfo{volume}{70}},
  \bibinfo{pages}{046608} (\bibinfo{year}{2004}).

\bibitem[{\citenamefont{Liu et~al.}(2008)\citenamefont{Liu, Cheng, Hand, Mock,
  Cui, Cummer, and Smith}}]{Liu}
\bibinfo{author}{\bibfnamefont{R.}~\bibnamefont{Liu}},
  \bibinfo{author}{\bibfnamefont{Q.}~\bibnamefont{Cheng}},
  \bibinfo{author}{\bibfnamefont{T.}~\bibnamefont{Hand}},
  \bibinfo{author}{\bibfnamefont{J.~J.} \bibnamefont{Mock}},
  \bibinfo{author}{\bibfnamefont{T.~J.} \bibnamefont{Cui}},
  \bibinfo{author}{\bibfnamefont{S.~A.} \bibnamefont{Cummer}},
  \bibnamefont{and} \bibinfo{author}{\bibfnamefont{D.~R.} \bibnamefont{Smith}},
  \bibinfo{journal}{Phys. Rev. Lett.} \textbf{\bibinfo{volume}{100}},
  \bibinfo{pages}{023903} (\bibinfo{year}{2008}).

\bibitem[{\citenamefont{Silveirinha and Engheta}(2006)}]{Silv}
\bibinfo{author}{\bibfnamefont{M.}~\bibnamefont{Silveirinha}} \bibnamefont{and}
  \bibinfo{author}{\bibfnamefont{N.}~\bibnamefont{Engheta}},
  \bibinfo{journal}{Phys. Rev. Lett.} \textbf{\bibinfo{volume}{97}},
  \bibinfo{pages}{157403} (\bibinfo{year}{2006}).

\bibitem[{\citenamefont{Edwards et~al.}(2008)\citenamefont{Edwards, Alu, Young,
  Silveirinha, and Engheta}}]{Bed1}
\bibinfo{author}{\bibfnamefont{B.}~\bibnamefont{Edwards}},
  \bibinfo{author}{\bibfnamefont{A.}~\bibnamefont{Alu}},
  \bibinfo{author}{\bibfnamefont{M.~E.} \bibnamefont{Young}},
  \bibinfo{author}{\bibfnamefont{M.}~\bibnamefont{Silveirinha}},
  \bibnamefont{and} \bibinfo{author}{\bibfnamefont{N.}~\bibnamefont{Engheta}},
  \bibinfo{journal}{Phys. Rev. Lett.} \textbf{\bibinfo{volume}{100}},
  \bibinfo{pages}{033903} (\bibinfo{year}{2008}).

\bibitem[{\citenamefont{Edwards et~al.}(2009)\citenamefont{Edwards, Alu,
  Silveirinha, and Engheta}}]{Bed2}
\bibinfo{author}{\bibfnamefont{B.}~\bibnamefont{Edwards}},
  \bibinfo{author}{\bibfnamefont{A.}~\bibnamefont{Alu}},
  \bibinfo{author}{\bibfnamefont{M.~G.} \bibnamefont{Silveirinha}},
  \bibnamefont{and} \bibinfo{author}{\bibfnamefont{N.}~\bibnamefont{Engheta}},
  \bibinfo{journal}{J. Appl. Phys.} \textbf{\bibinfo{volume}{105}},
  \bibinfo{pages}{044905} (\bibinfo{year}{2009}).

\bibitem[{\citenamefont{Liu et~al.}(2009)\citenamefont{Liu, Hu, Zhao, and
  Luo}}]{Luo}
\bibinfo{author}{\bibfnamefont{L.}~\bibnamefont{Liu}},
  \bibinfo{author}{\bibfnamefont{C.}~\bibnamefont{Hu}},
  \bibinfo{author}{\bibfnamefont{Z.}~\bibnamefont{Zhao}}, \bibnamefont{and}
  \bibinfo{author}{\bibfnamefont{X.}~\bibnamefont{Luo}}, \bibinfo{journal}{Opt.
  Exp.} \textbf{\bibinfo{volume}{17}}, \bibinfo{pages}{12183}
  (\bibinfo{year}{2009}).

\bibitem[{\citenamefont{Pollard et~al.}(2009)\citenamefont{Pollard, Murphy,
  Hendren, Evans, Atkinson, Wurtz, Zayats, and Podolskiy}}]{Pollard}
\bibinfo{author}{\bibfnamefont{R.~J.} \bibnamefont{Pollard}},
  \bibinfo{author}{\bibfnamefont{A.}~\bibnamefont{Murphy}},
  \bibinfo{author}{\bibfnamefont{W.~R.} \bibnamefont{Hendren}},
  \bibinfo{author}{\bibfnamefont{P.~R.} \bibnamefont{Evans}},
  \bibinfo{author}{\bibfnamefont{R.}~\bibnamefont{Atkinson}},
  \bibinfo{author}{\bibfnamefont{G.~A.} \bibnamefont{Wurtz}},
  \bibinfo{author}{\bibfnamefont{A.~V.} \bibnamefont{Zayats}},
  \bibnamefont{and} \bibinfo{author}{\bibfnamefont{V.~A.}
  \bibnamefont{Podolskiy}}, \bibinfo{journal}{Phys. Rev. Lett.}
  \textbf{\bibinfo{volume}{102}}, \bibinfo{pages}{127405}
  (\bibinfo{year}{2009}).

\bibitem[{\citenamefont{Li et~al.}(1993)\citenamefont{Li, Stradling, Knight,
  Birch, Thomas, Phillips, and Ferguson}}]{Li}
\bibinfo{author}{\bibfnamefont{Y.~B.} \bibnamefont{Li}},
  \bibinfo{author}{\bibfnamefont{R.~A.} \bibnamefont{Stradling}},
  \bibinfo{author}{\bibfnamefont{T.}~\bibnamefont{Knight}},
  \bibinfo{author}{\bibfnamefont{J.~R.} \bibnamefont{Birch}},
  \bibinfo{author}{\bibfnamefont{R.~H.} \bibnamefont{Thomas}},
  \bibinfo{author}{\bibfnamefont{C.~C.} \bibnamefont{Phillips}},
  \bibnamefont{and} \bibinfo{author}{\bibfnamefont{I.~T.}
  \bibnamefont{Ferguson}}, \bibinfo{journal}{Semicond. Sci. Technol.}
  \textbf{\bibinfo{volume}{8}}, \bibinfo{pages}{101} (\bibinfo{year}{1993}).

\bibitem[{\citenamefont{Alu et~al.}(2006)\citenamefont{Alu, Bilotti, Engheta,
  and Vegni}}]{Alu}
\bibinfo{author}{\bibfnamefont{A.}~\bibnamefont{Alu}},
  \bibinfo{author}{\bibfnamefont{F.}~\bibnamefont{Bilotti}},
  \bibinfo{author}{\bibfnamefont{N.}~\bibnamefont{Engheta}}, \bibnamefont{and}
  \bibinfo{author}{\bibfnamefont{L.}~\bibnamefont{Vegni}},
  \bibinfo{journal}{IEEE Trans. Antennas Propag.}
  \textbf{\bibinfo{volume}{54}}, \bibinfo{pages}{1632} (\bibinfo{year}{2006}).

\end{thebibliography}


\begin{thebibliography}{27}
\expandafter\ifx\csname natexlab\endcsname\relax\def\natexlab#1{#1}\fi
\expandafter\ifx\csname bibnamefont\endcsname\relax
  \def\bibnamefont#1{#1}\fi
\expandafter\ifx\csname bibfnamefont\endcsname\relax
  \def\bibfnamefont#1{#1}\fi
\expandafter\ifx\csname citenamefont\endcsname\relax
  \def\citenamefont#1{#1}\fi
\expandafter\ifx\csname url\endcsname\relax
  \def\url#1{\texttt{#1}}\fi
\expandafter\ifx\csname urlprefix\endcsname\relax\def\urlprefix{URL }\fi
\providecommand{\bibinfo}[2]{#2}
\providecommand{\eprint}[2][]{\url{#2}}

\bibitem[{\citenamefont{Merlin}(2007)}]{Merlin}
\bibinfo{author}{\bibfnamefont{R.}~\bibnamefont{Merlin}},
  \bibinfo{journal}{Science} \textbf{\bibinfo{volume}{317}},
  \bibinfo{pages}{927} (\bibinfo{year}{2007}).

\bibitem[{\citenamefont{Markley et~al.}(2008)\citenamefont{Markley, Wong, Wang,
  and Eleftheriades}}]{Eleftheriades}
\bibinfo{author}{\bibfnamefont{L.}~\bibnamefont{Markley}},
  \bibinfo{author}{\bibfnamefont{A.}~\bibnamefont{Wong}},
  \bibinfo{author}{\bibfnamefont{Y.}~\bibnamefont{Wang}}, \bibnamefont{and}
  \bibinfo{author}{\bibfnamefont{G.}~\bibnamefont{Eleftheriades}},
  \bibinfo{journal}{Phys. Rev. Lett.} \textbf{\bibinfo{volume}{101}},
  \bibinfo{pages}{113901} (\bibinfo{year}{2008}).

\bibitem[{\citenamefont{Thongrattanasiri and Podolskiy}(2009)}]{Podolskiy1}
\bibinfo{author}{\bibfnamefont{S.}~\bibnamefont{Thongrattanasiri}}
  \bibnamefont{and}
  \bibinfo{author}{\bibfnamefont{V.}~\bibnamefont{Podolskiy}},
  \bibinfo{journal}{Opt. Lett.} \textbf{\bibinfo{volume}{34}},
  \bibinfo{pages}{890} (\bibinfo{year}{2009}).

\bibitem[{\citenamefont{Pendry}(2000)}]{superlens}
\bibinfo{author}{\bibfnamefont{J.}~\bibnamefont{Pendry}},
  \bibinfo{journal}{Phys. Rev. Lett.} \textbf{\bibinfo{volume}{85}},
  \bibinfo{pages}{3966} (\bibinfo{year}{2000}).

\bibitem[{\citenamefont{Liu et~al.}(2007)\citenamefont{Liu, Lee, Xiong, Sun,
  and Zhang}}]{Zhang1}
\bibinfo{author}{\bibfnamefont{Z.}~\bibnamefont{Liu}},
  \bibinfo{author}{\bibfnamefont{H.}~\bibnamefont{Lee}},
  \bibinfo{author}{\bibfnamefont{Y.}~\bibnamefont{Xiong}},
  \bibinfo{author}{\bibfnamefont{C.}~\bibnamefont{Sun}}, \bibnamefont{and}
  \bibinfo{author}{\bibfnamefont{X.}~\bibnamefont{Zhang}},
  \bibinfo{journal}{Science} \textbf{\bibinfo{volume}{315}},
  \bibinfo{pages}{1686} (\bibinfo{year}{2007}).

\bibitem[{\citenamefont{Jacob et~al.}(2006)\citenamefont{Jacob, Alekseyev, and
  Narimanov}}]{hyperlensEN}
\bibinfo{author}{\bibfnamefont{Z.}~\bibnamefont{Jacob}},
  \bibinfo{author}{\bibfnamefont{L.}~\bibnamefont{Alekseyev}},
  \bibnamefont{and}
  \bibinfo{author}{\bibfnamefont{E.}~\bibnamefont{Narimanov}},
  \bibinfo{journal}{Opt.Exp.} \textbf{\bibinfo{volume}{14}},
  \bibinfo{pages}{8247} (\bibinfo{year}{2006}).

\bibitem[{\citenamefont{Salandrino and Engheta}(2006)}]{hyperlensNE}
\bibinfo{author}{\bibfnamefont{A.}~\bibnamefont{Salandrino}} \bibnamefont{and}
  \bibinfo{author}{\bibfnamefont{N.}~\bibnamefont{Engheta}},
  \bibinfo{journal}{Phys.Rev.B} \textbf{\bibinfo{volume}{74}},
  \bibinfo{pages}{075103} (\bibinfo{year}{2006}).

\bibitem[{\citenamefont{Smolyaninov et~al.}(2007)\citenamefont{Smolyaninov,
  Hung, and Davis}}]{hyperlensIS}
\bibinfo{author}{\bibfnamefont{I.}~\bibnamefont{Smolyaninov}},
  \bibinfo{author}{\bibfnamefont{Y.}~\bibnamefont{Hung}}, \bibnamefont{and}
  \bibinfo{author}{\bibfnamefont{C.}~\bibnamefont{Davis}},
  \bibinfo{journal}{Science} \textbf{\bibinfo{volume}{315}},
  \bibinfo{pages}{1699} (\bibinfo{year}{2007}).

\bibitem[{\citenamefont{Ziolkowski}(2004)}]{Ziol}
\bibinfo{author}{\bibfnamefont{R.}~\bibnamefont{Ziolkowski}},
  \bibinfo{journal}{Phys. Rev. E} \textbf{\bibinfo{volume}{70}},
  \bibinfo{pages}{046608} (\bibinfo{year}{2004}).

\bibitem[{\citenamefont{Silveirinha and Engheta}(2006)}]{Engheta1}
\bibinfo{author}{\bibfnamefont{M.}~\bibnamefont{Silveirinha}} \bibnamefont{and}
  \bibinfo{author}{\bibfnamefont{N.}~\bibnamefont{Engheta}},
  \bibinfo{journal}{Phys. Rev. Lett.} \textbf{\bibinfo{volume}{97}},
  \bibinfo{pages}{157403} (\bibinfo{year}{2006}).

\bibitem[{\citenamefont{Liu et~al.}(2008)\citenamefont{Liu, Cheng, Hand, Mock,
  Cui, Cummer, and Smith}}]{DSmith1}
\bibinfo{author}{\bibfnamefont{R.}~\bibnamefont{Liu}},
  \bibinfo{author}{\bibfnamefont{Q.}~\bibnamefont{Cheng}},
  \bibinfo{author}{\bibfnamefont{T.}~\bibnamefont{Hand}},
  \bibinfo{author}{\bibfnamefont{J.}~\bibnamefont{Mock}},
  \bibinfo{author}{\bibfnamefont{T.}~\bibnamefont{Cui}},
  \bibinfo{author}{\bibfnamefont{S.}~\bibnamefont{Cummer}}, \bibnamefont{and}
  \bibinfo{author}{\bibfnamefont{D.}~\bibnamefont{Smith}},
  \bibinfo{journal}{Phys. Rev. Lett.} \textbf{\bibinfo{volume}{100}},
  \bibinfo{pages}{023903} (\bibinfo{year}{2008}).

\bibitem[{\citenamefont{Edwards et~al.}(2008)\citenamefont{Edwards, Alu, Young,
  Silveirinha, and Engheta}}]{Engheta2}
\bibinfo{author}{\bibfnamefont{B.}~\bibnamefont{Edwards}},
  \bibinfo{author}{\bibfnamefont{A.}~\bibnamefont{Alu}},
  \bibinfo{author}{\bibfnamefont{M.}~\bibnamefont{Young}},
  \bibinfo{author}{\bibfnamefont{M.}~\bibnamefont{Silveirinha}},
  \bibnamefont{and} \bibinfo{author}{\bibfnamefont{N.}~\bibnamefont{Engheta}},
  \bibinfo{journal}{Phys. Rev. Lett.} \textbf{\bibinfo{volume}{100}},
  \bibinfo{pages}{033903} (\bibinfo{year}{2008}).

\bibitem[{\citenamefont{Alu et~al.}(2006)\citenamefont{Alu, Bilotti, Engheta,
  and Vegni}}]{Engheta3}
\bibinfo{author}{\bibfnamefont{A.}~\bibnamefont{Alu}},
  \bibinfo{author}{\bibfnamefont{F.}~\bibnamefont{Bilotti}},
  \bibinfo{author}{\bibfnamefont{N.}~\bibnamefont{Engheta}}, \bibnamefont{and}
  \bibinfo{author}{\bibfnamefont{L.}~\bibnamefont{Vegni}},
  \bibinfo{journal}{IEEE Trans. Antennas Propag.}
  \textbf{\bibinfo{volume}{54}}, \bibinfo{pages}{1632} (\bibinfo{year}{2006}).

\bibitem[{\citenamefont{Boltasseva and Atwater}(2011)}]{Bolt}
\bibinfo{author}{\bibfnamefont{A.}~\bibnamefont{Boltasseva}} \bibnamefont{and}
  \bibinfo{author}{\bibfnamefont{H.}~\bibnamefont{Atwater}},
  \bibinfo{journal}{Science} \textbf{\bibinfo{volume}{331}},
  \bibinfo{pages}{290} (\bibinfo{year}{2011}).

\bibitem[{\citenamefont{Thongrattanasiri
  et~al.}(2009)\citenamefont{Thongrattanasiri, Elser, and
  Podolskiy}}]{sukoModeMatch}
\bibinfo{author}{\bibfnamefont{S.}~\bibnamefont{Thongrattanasiri}},
  \bibinfo{author}{\bibfnamefont{J.}~\bibnamefont{Elser}}, \bibnamefont{and}
  \bibinfo{author}{\bibfnamefont{V.}~\bibnamefont{Podolskiy}},
  \bibinfo{journal}{J. Opt. Soc. Am. B} \textbf{\bibinfo{volume}{26}},
  \bibinfo{pages}{B102} (\bibinfo{year}{2009}).

\bibitem[{\citenamefont{Jin et~al.}(2010)\citenamefont{Jin, Zhang, and
  He}}]{He}
\bibinfo{author}{\bibfnamefont{Y.}~\bibnamefont{Jin}},
  \bibinfo{author}{\bibfnamefont{P.}~\bibnamefont{Zhang}}, \bibnamefont{and}
  \bibinfo{author}{\bibfnamefont{S.}~\bibnamefont{He}}, \bibinfo{journal}{Phys.
  Rev. B} \textbf{\bibinfo{volume}{82}}, \bibinfo{pages}{075118}
  (\bibinfo{year}{2010}).

\bibitem[{\citenamefont{Ebbesen et~al.}(1998)\citenamefont{Ebbesen, Lezec,
  Ghaemi, Thio, and Wolff}}]{EOT}
\bibinfo{author}{\bibfnamefont{T.~W.} \bibnamefont{Ebbesen}},
  \bibinfo{author}{\bibfnamefont{H.~J.} \bibnamefont{Lezec}},
  \bibinfo{author}{\bibfnamefont{H.~F.} \bibnamefont{Ghaemi}},
  \bibinfo{author}{\bibfnamefont{T.}~\bibnamefont{Thio}}, \bibnamefont{and}
  \bibinfo{author}{\bibfnamefont{P.~A.} \bibnamefont{Wolff}},
  \bibinfo{journal}{Nature} \textbf{\bibinfo{volume}{391}},
  \bibinfo{pages}{667} (\bibinfo{year}{1998}).

\bibitem[{\citenamefont{Lezec et~al.}(2002)\citenamefont{Lezec, Degiron,
  Devaux, Linke, Martin-Moreno, Garcia-Vidal, and Ebbesen}}]{beamsteer}
\bibinfo{author}{\bibfnamefont{H.}~\bibnamefont{Lezec}},
  \bibinfo{author}{\bibfnamefont{A.}~\bibnamefont{Degiron}},
  \bibinfo{author}{\bibfnamefont{E.}~\bibnamefont{Devaux}},
  \bibinfo{author}{\bibfnamefont{R.}~\bibnamefont{Linke}},
  \bibinfo{author}{\bibfnamefont{L.}~\bibnamefont{Martin-Moreno}},
  \bibinfo{author}{\bibfnamefont{F.}~\bibnamefont{Garcia-Vidal}},
  \bibnamefont{and} \bibinfo{author}{\bibfnamefont{T.}~\bibnamefont{Ebbesen}},
  \bibinfo{journal}{Science} \textbf{\bibinfo{volume}{297}},
  \bibinfo{pages}{820} (\bibinfo{year}{2002}).

\bibitem[{\citenamefont{Pendry et~al.}(2006)\citenamefont{Pendry, Schurig, and
  Smith}}]{cloakJP}
\bibinfo{author}{\bibfnamefont{J.}~\bibnamefont{Pendry}},
  \bibinfo{author}{\bibfnamefont{D.}~\bibnamefont{Schurig}}, \bibnamefont{and}
  \bibinfo{author}{\bibfnamefont{D.}~\bibnamefont{Smith}},
  \bibinfo{journal}{Science} \textbf{\bibinfo{volume}{312}},
  \bibinfo{pages}{1780} (\bibinfo{year}{2006}).

\bibitem[{\citenamefont{Cai et~al.}(2007)\citenamefont{Cai, Chettiar,
  Kildishev, and Shalaev}}]{cloakVS}
\bibinfo{author}{\bibfnamefont{W.}~\bibnamefont{Cai}},
  \bibinfo{author}{\bibfnamefont{U.}~\bibnamefont{Chettiar}},
  \bibinfo{author}{\bibfnamefont{A.}~\bibnamefont{Kildishev}},
  \bibnamefont{and} \bibinfo{author}{\bibfnamefont{V.}~\bibnamefont{Shalaev}},
  \bibinfo{journal}{Nat. Phot.} \textbf{\bibinfo{volume}{1}},
  \bibinfo{pages}{224} (\bibinfo{year}{2007}).

\bibitem[{\citenamefont{Kildishev and Shalaev}(2008)}]{blackHoleVS}
\bibinfo{author}{\bibfnamefont{A.}~\bibnamefont{Kildishev}} \bibnamefont{and}
  \bibinfo{author}{\bibfnamefont{V.}~\bibnamefont{Shalaev}},
  \bibinfo{journal}{Opt.Lett} \textbf{\bibinfo{volume}{33}},
  \bibinfo{pages}{43} (\bibinfo{year}{2008}).

\bibitem[{\citenamefont{Engheta}(2007)}]{enghetaCircuits}
\bibinfo{author}{\bibfnamefont{N.}~\bibnamefont{Engheta}},
  \bibinfo{journal}{Science} \textbf{\bibinfo{volume}{317}},
  \bibinfo{pages}{1698} (\bibinfo{year}{2007}).

\bibitem[{\citenamefont{Li et~al.}(1993)\citenamefont{Li, Stradling, Knight,
  Birch, Thomas, Phillips, and Ferguson}}]{Li}
\bibinfo{author}{\bibfnamefont{Y.~B.} \bibnamefont{Li}},
  \bibinfo{author}{\bibfnamefont{R.~A.} \bibnamefont{Stradling}},
  \bibinfo{author}{\bibfnamefont{T.}~\bibnamefont{Knight}},
  \bibinfo{author}{\bibfnamefont{J.~R.} \bibnamefont{Birch}},
  \bibinfo{author}{\bibfnamefont{R.~H.} \bibnamefont{Thomas}},
  \bibinfo{author}{\bibfnamefont{C.~C.} \bibnamefont{Phillips}},
  \bibnamefont{and} \bibinfo{author}{\bibfnamefont{I.~T.}
  \bibnamefont{Ferguson}}, \bibinfo{journal}{Semicond. Sci. Technol.}
  \textbf{\bibinfo{volume}{8}}, \bibinfo{pages}{101} (\bibinfo{year}{1993}).

\bibitem[{\citenamefont{Born and Wolf}(1999)}]{bornWolf}
\bibinfo{author}{\bibfnamefont{M.}~\bibnamefont{Born}} \bibnamefont{and}
  \bibinfo{author}{\bibfnamefont{E.}~\bibnamefont{Wolf}},
  \emph{\bibinfo{title}{Principles of Optics}} (\bibinfo{publisher}{Cambridge
  U. Press}, \bibinfo{year}{1999}).

\bibitem[{\citenamefont{Schevchenko}(1971)}]{schevchenkoBook}
\bibinfo{author}{\bibfnamefont{V.~V.} \bibnamefont{Schevchenko}},
  \emph{\bibinfo{title}{Continuous Transitions in Open Waveguides}}
  (\bibinfo{publisher}{Golem}, \bibinfo{year}{1971}).

\bibitem[{\citenamefont{Bravo-Abad et~al.}(2004)\citenamefont{Bravo-Abad,
  Martin-Moreno, and Garcia-Vidal}}]{SingleSlit}
\bibinfo{author}{\bibfnamefont{J.}~\bibnamefont{Bravo-Abad}},
  \bibinfo{author}{\bibfnamefont{L.}~\bibnamefont{Martin-Moreno}},
  \bibnamefont{and}
  \bibinfo{author}{\bibfnamefont{F.}~\bibnamefont{Garcia-Vidal}},
  \bibinfo{journal}{Phys. Rev. E} \textbf{\bibinfo{volume}{69}},
  \bibinfo{pages}{026601} (\bibinfo{year}{2004}).

\bibitem[{com()}]{comsol}
\bibinfo{note}{Www.comsol.com}.

\end{thebibliography}

\end{document}